\documentstyle[12pt,epsf,aasms4]{article}
\input{psfig.sty}
\begin{document}

\title{Convective instability in proto--neutron stars}

\author{Juan A. Miralles$^{1)}$, Jos\'e A. Pons$^{1,2)}$ and Vadim A. Urpin$^{3,4)}$} 

\affil{$^{1)}$ Departament d'Astronomia i Astrof\'{\i}sica, Universitat de
                    Val\`encia, E-46100 Burjassot, Spain \\
$^{2)}$ Department of Physics \& Astronomy, SUNY at Stony Brook,
Stony Brook, NY 11794-3800, USA \\
$^{3)}$ Department of Mathematics, University of Newcastle, 
                   Newcastle upon Tyne NE1 7RU, UK \\
$^{4)}$ A.F.Ioffe Institute of Physics and Technology, 
                   194021 St. Petersburg, Russia}
 
\begin{abstract}

The linear hydrodynamic stability of proto--neutron stars (PNSs) is considered
taking into account dissipative processes such as neutrino transport 
and viscosity. We obtain the general
instability criteria which differ essentially from the well-known
Ledoux criterion used in previous studies. 
We apply the criteria to evolutive models of PNSs that,
in general, can be subject to the various known regimes such as neutron 
fingers 
and convective instabilities. Our results indicate that
the fingers instability arises in a more extended region of
the stellar volume and lasts a longer time than expected.

\end{abstract}

\vspace{1cm}

\newpage

\section{Introduction}

According to theoretical models, neutron stars are formed in the aftermath 
of type II and Ib supernova explosions, associated to the gravitational
collapse of massive stars (8 to 30$M_{\odot}$) at the end of their lives.
A number of authors have recognized 
that convection in the newly born hot neutron star can play an important 
role in both enhancing neutrino luminosities and increasing the energy
deposition efficiency, which might lead to the explosion of a supernova 
(Epstein 1979; Livio, Buchler \& Colgate 1980; Colgate \& Petschek 1980; 
Smarr {\it et al.}, 1981; Lattimer \& Mazurek, 1981; 
Burrows\& Fryxell 1993; Janka \& Muller 1994).
Epstein (1979) pointed out that the negative lepton gradient that
naturally arises in the outer layers of the PNS after the 
shock breaks out of the neutrinosphere can be convectively unstable. 
Later on, Colgate
\& Petschek (1980) suggested that this instability could lead to a complete
overturn of the core and a strong enhancement of the neutrino transport needed
for a powerful supernova explosion on a dynamical time scale. However,
convection in PNSs can be driven not only by the lepton 
gradient but the entropy gradient as well.
The suggestion that dissipation 
of the shock must result in a negative entropy gradient due to both 
neutronization and dissociation (Arnett, 1987) has been confirmed in a 
variety of models by Burrows \& Lattimer (1988). As a matter of fact, 
despite different equations of state and differences in the consideration 
of neutrino transport, the development of negative entropy and lepton 
gradients seems to be common in many simulations of supernova models 
(Hillebrandt 1987, Bruenn \& Mezzacappa 1994, Bruenn, Mezzacappa \& Dineva 
1995) and evolutionary models of PNSs (Burrows \& Lattimer 
1986, Keil \& Janka 1995, Sumiyoshi, Suzuki \& Toki 1995, Pons et al. 
1999). 

Concerning numerical simulations, the situation is controversial.
Recent hydrodynamic simulations by Keil, Janka \& Muller (1996) 
demonstrate that convection can arise in a rather extended region of 
PNSs and may generally last a relatively long time, $>1$s. 
Bruenn and Mezzacappa (1994), using the mixing length approximation and 
Mezzacappa et. al (1998) in two dimensional hydrodynamic simulations 
found only mild convective activity in the region near the neutrinosphere.
Note that the two-dimensional hydro code used by Keil, Janka \& Muller 
is coupled to a radial gray 
equilibrium diffusion code, that suppresses the neutrino transport in the 
angular direction, essentially underestimating the stabilizing effect of 
neutrino transport. On the contrary, Mezzacappa et. al (1998) code works 
at the limit in which the neutrino transport in the angular direction is 
fast enough to render the neutrino distributions spherically symmetric, 
therefore overestimating the stabilizing effect. 
The answer is obviously at some point in between, that only a future 
self--consistent multidimensional calculation can determine.

Helpful insights into the nature and growth rates of fluid instabilities
have been achieved thanks to 
semi-analytical investigations (Grossman, Narayan 
\& Arnett, 1993; Bruenn \& Dineva 1996), and this work is in that line.
The presence of convection in PNSs is usually argued by the 
fact that the necessary condition of instability (the Ledoux criterion) 
is fulfilled in a some fraction of the stellar volume.
(Epstein 1979, Livio; Buchler \& Colgate  1980; 
Keil, Janka \& M\"uller, 1996). However,
the Ledoux criterion may have no bearing at all on the physics of convection. 
If dissipative effects caused by viscosity or
energy and lepton transport are taken into account, the Ledoux condition
does not apply and has to be substituted by the appropriate criteria.
In the present paper, we derived the criteria of instability in 
PNSs employing the diffusion approximation for neutrino 
transport (Imshennik \& Nadezhin 1972, Pons et al. 1999) and show that 
dissipative processes can substantially change the picture.

\section{Basic equations}

Consider the condition of instability in a plane-parallel layer 
between $z=0$ and $z=d$ with the gravity ${\bf g}$ directed in the negative 
$z$-direction. We neglect a non-uniformity of ${\bf g}$ as well as general 
relativistic corrections to hydrodynamic equations. Calculations show
that initially the instability arises in a surface layer of PNS thus
the plane-parallel approximation seems to be accurate, at least, during
the initial evolutionary stage and, should give qualitatively correct 
results during the late stage. We assume that the 
characteristic cooling time scale of a PNS is much longer 
than the growth time of instability thus it can be treated in a 
quasi-stationary approximation. Since convective velocities are 
typically smaller than the speed of sound, one can describe instability
by making use of the standard Boussinesq approximation (see, e.g., Landau 
\& Lifshitz 1959). We consider the linear instability when the equations 
governing small perturbations can be obtained by the linearization of 
hydrodynamic equations. In what follows, small perturbations of hydrodynamic 
quantities will be marked by a subscript ``1''. The linearized momentum and 
continuity equations read
\begin{equation}
\rho \dot{\bf v}_{1} = - \nabla p_{1} + {\bf g} \rho_{1}+ \rho \nu \Delta 
{\bf v}_{1} ,
\label{hyd1}
\end{equation}
\begin{equation}
\nabla \cdot  {\bf v}_{1} = 0,
\label{hyd0}
\end{equation}
where $p$ and $\rho$ are the pressure and density, respectively, $\nu$ is
the kinematic viscosity. We assume that the matter inside a PNS 
is in chemical equilibrium thus the density is generally a function of 
the pressure $p$, temperature $T$ and lepton fraction $Y=(n_{e} + 
n_{\nu})/n$, with $n_{e}$ and $n_{\nu}$ being the net (particles minus 
antiparticles) number densities of electrons and neutrinos, respectively, and
$n=n_{p}+n_{n}$ is the number density of baryons. 
Since in the Boussinesq approximation, the perturbations of pressure are
negligible because the fluid motions are assumed to be slow and the
moving fluid elements are nearly in pressure equilibrium with surroundings,  
the perturbations of density ($\rho_1$) and entropy per baryon ($s_1$) can
be expressed, in terms of the perturbations of temperature ($T_1$) and 
lepton fraction $Y_1$,
\begin{equation}
\rho_{1} \approx - \rho ( \beta \frac{T_{1}}{T} + \delta Y_{1} ),
\label{rho1}
\end{equation}
\begin{equation}
s_1\approx {m_B c_{p}} \frac{T_{1}}{T} + \sigma Y_{1}
\label{t1}
\end{equation} 
where $m_B$ is the mass of the baryon (we neglect the mass difference
between protons and neutrons), $\beta$ and $\delta$ are the coefficients 
of thermal and chemical
expansion; $\beta = - (\partial \ln \rho/\partial \ln T)_{pY}$, $\delta =
- (\partial \ln \rho/ \partial Y)_{pT}$, $c_{p} = (T/m_B) 
(\partial s / \partial T)_{pY}$ is the specific heat at
constant pressure and $\sigma = (\partial s/ \partial Y)_{pT}$.

The above equations should be complemented by the equation driving
the evolution of chemical composition and heat balance. We employ
the equilibrium diffusion approximation (EDA) which can be reliable  
during the early stage of evolution of the PNS, when the mean 
free path of neutrino is short compared to the density and temperature 
length scales. 
In this approximation, the diffusion and thermal balance equations
read 
\begin{equation}
n \frac{d Y}{d t} = - \nabla \cdot {\bf F},
\label{yev}
\end{equation}
\begin{equation}
n T \frac{d s}{d t} + n \mu \frac{d Y}{d t}= - \nabla \cdot {\bf H},
\label{sev}
\end{equation} 
where $\mu$ is the neutrino chemical potential and 
${\bf F}$ and ${\bf H}$ are the lepton and heat fluxes, respectively. 
Note that in the equation for the heat balance we have neglected viscous 
heat production since this term is second order in velocity. In the EDA, 
these fluxes are given by
\begin{equation}
{\bf F} = - a_{T} \nabla T - a_{\eta} \nabla \eta,
\end{equation}
\begin{equation}
{\bf H} = - b_{T} \nabla T - b_{\eta} \nabla \eta,
\end{equation}
where $\eta= \mu/k_B T$ is the degeneracy parameter of the neutrino
gas, $k_B$ being the Boltzmann constant; $a_{T}$, $a_{\eta}$ and $b_{T}$, 
$b_{\eta}$ are the coefficients of diffusion and the thermal conductivities,
respectively.
These kinetic coefficients can easily be related to the coefficients 
$D_{2}$, $D_{3}$ and $D_{4}$ introduced by Pons et al. (1999),
$a_{T}=\Gamma D_{3}$, $a_{\eta}= T \Gamma D_{2}$, $b_{T}=k_B T \Gamma D_{4}$, 
$b_{\eta}= k_B T^{2} \Gamma D_{3}$, where 
$\Gamma = (k_B T)^{2} k_B/ 6 \pi^{2} c^{2} \hbar^{3}$. 

For hydrodynamic considerations, it is more convenient to express ${\bf F}$
and ${\bf H}$ in terms of $\nabla T$ and $\nabla Y$ rather
than $\nabla T$ and $\nabla \eta$. Then, neglecting the contribution
of the pressure gradient (barodiffusion), we have
\begin{equation}
{\bf F} = - \left( \chi_{T} \nabla T +  \chi_{Y} \nabla Y \right),
\label{f1}
\end{equation}
\begin{equation}
{\bf H} = - (\xi_{T} \nabla T + \xi_{Y} \nabla Y),
\label{f2}
\end{equation}
where 
\begin{eqnarray}
\chi_{T} = a_{T}+ a_{\eta} \left( 
\frac{\partial \eta}{\partial T} \right)_{p,Y} \, , \;\; \chi_{Y} = 
a_{\eta} \left( \frac{\partial \eta}{\partial Y} \right)_{p,T} \, ,
\nonumber \\
\xi_{T} = b_{T} + b_{\eta} \left( 
\frac{\partial \eta}{\partial T} \right)_{p,Y} \, , \;\; \xi_{Y} =  
b_{\eta} \left( \frac{\partial \eta}{\partial Y} \right)_{p,T} \, .
\end{eqnarray}
Assuming that 
unperturbed quantities are homogeneous, equations (\ref{yev}) and
(\ref{sev}) can be linearized and written in the form
\begin{eqnarray}
\dot{Y}_{1} &+& {\bf v}_{1} \cdot \nabla Y =  
\lambda_{T} \frac{\Delta T_{1}}{T} + \lambda_{Y} \Delta Y_{1}\;,
\label{y1}
\\
\frac{\dot{T}_{1}}{T} &-& {\bf v}_{1} \cdot  \frac{\Delta\nabla T}{T} =
\kappa_{T} \frac{\Delta T_{1}}{T} + \kappa_{Y} \Delta Y_{1} 
\label{s1}
\end{eqnarray}
where 
\begin{equation}
\Delta\nabla T= 
-\frac{T}{m_B c_p}  \left( \nabla s - \sigma \nabla Y  \right)
=  \left( \frac{\partial T}{\partial p} \right)_{s,Y} \nabla p
-  \nabla T 
\end{equation}
is the superadiabatic 
temperature gradient that gives the difference between the temperature 
gradient of a fluid with constant entropy and composition and the 
actual temperature gradient,
and we have introduced the following characteristic conductivities
\begin{eqnarray}
\kappa_{T} = \frac{1}{m_B c_p} \left[
\frac{\xi_{T}}{n} - \lambda_T ( \eta + \sigma ) \right], &\;\;&
\lambda_{T} = \frac{T \chi_{T}}{n},
\nonumber \\
\kappa_{Y} = \frac{1}{m_B c_p} \left[
\frac{\xi_{Y}}{n T} - \lambda_{Y} ( \eta + \sigma ) \right], &\;\;&
\lambda_{Y} = \frac{\chi_{Y}}{n}.
\end{eqnarray}

The system formed by equations (\ref{hyd1}), (\ref{hyd0}), (\ref{y1}) and 
(\ref{s1}), together with the corresponding boundary conditions, completely 
determines the behaviour of small perturbations. For the sake of 
simplicity, we consider the case when perturbations are vanishing at the 
boundaries $z=0$ and $z=d$. Note that other boundary conditions cannot 
change the main conclusion of our paper qualitatively.

\section{The dispersion equation}

The dependence of all perturbations on time and horizontal coordinate 
can be chosen in the form $\exp(\gamma t - i k x)$, where $\gamma$ 
is the inverse growth (or decay) timescale of perturbations and $k$ is the 
horizontal wavevector. For such perturbations, we have
\begin{equation}
\Delta = \frac{d^{2}}{d z^{2}} - k^{2}.
\end{equation}
The dependence on the vertical coordinate should be obtained from 
equations (\ref{hyd1}), (\ref{hyd0}), (\ref{y1}) and (\ref{s1}), which 
can be reduced to one equation of a higher order, say for $v_{1z}$. 
The coefficients of this equation are constant in our simplified model,
therefore the solution for the fundamental mode with ``zero boundary
conditions'' can be taken in the form $v_{1z} = \sin (\pi z/ d)$. Then,
the dispersion equation for the fundamental mode is
\begin{equation}
\gamma^{3} + a_{2} \gamma^{2} + a_{1} \gamma + a_{0} = 0,
\label{cubic}
\end{equation}
where 
\begin{eqnarray}
a_{2} = \omega_{\nu} + \omega_{T} + \omega_{Y},  \nonumber  \\
a_{1} = \omega_{T} \omega_{Y} - \omega_{TY} \omega_{YT} +
\omega_{\nu} (\omega_{T} + \omega_{Y}) - (\omega_{g}^{2} +
\omega_{L}^{2}),  \nonumber  \\
a_{0} = \omega_{\nu} (\omega_{T} \omega_{Y} - \omega_{TY} \omega_{YT})
- \omega_{g}^{2} ( \omega_{Y} - \frac{\delta}{\beta} \omega_{YT}) 
\nonumber  \\
- \omega_{L}^{2} \left( \omega_{T} - \frac{\beta}{\delta} \omega_{TY} \right).
\end{eqnarray}
In these expressions, we introduced the characteristic frequencies
\begin{eqnarray}
\omega_{T} = \kappa_{T} Q^{2}, &\;& 
\omega_{Y} = \lambda_{Y} Q^{2},   
\nonumber \\
\omega_{YT} = \lambda_{T} Q^{2}, &\;&
\omega_{TY} = \kappa_{Y} Q^{2}, \\
\omega_{\nu} = \nu Q^{2}, \quad \;
\omega_{g}^{2} = \frac{\beta g k^{2}}{Q^{2}} 
\cdot \frac{(\Delta\nabla T)_z}{T}, &\;&
\omega_{L}^{2} = - \frac{\delta g k^{2}}{Q^{2}} \cdot 
\frac{d Y}{d z},        \nonumber
\end{eqnarray}
where $(\Delta\nabla T)_z$ is the $z$-component of 
$\Delta\nabla T$ and $Q^{2} = (\pi/d)^{2} + k^{2}$. 
The quantities $\omega_{\nu}$, $\omega_{T}$, and $\omega_{Y}$ are the
inverse time scales of dissipation of perturbations due to viscosity,
thermal conductivity and diffusivity, respectively; $\omega_{YT}$
characterizes the rate of diffusion caused by the temperature 
inhomogeneity (thermodiffusion), and $\omega_{TY}$ describes the influence 
of the chemical inhomogeneity on the rate of heat conduction;
$\omega_{g}$ is the frequency (or, in the case of instability, the
inverse growth time) of the buoyant wave; $\omega_{L}$ characterizes 
the dynamical time scale of the processes associated with the lepton 
gradient.

Equation (\ref{cubic}) describes three essentially different modes which 
generally exist in a chemically inhomogeneous fluid. The condition that
at least one of the roots has a positive real part
(unstable) is equivalent to fulfilling one of the following inequalities
(see, e.g., Aleksandrov, Kolmogorov \& Laurentiev 1963)
\begin{equation}
a_{2} < 0, \;\;\; a_{0} < 0, \;\;\; a_{1} a_{2} < a_{0}.
\label{3cond}
\end{equation} 
Since $\nu$, $\kappa_T$ and $\lambda_Y$ are positive defined quantities, 
the first condition $a_2<0$ will never apply, and the discussion
will be reduced to the other two conditions.

In the particular case of chemically homogeneous plasma with a ``standard''
transport ($\omega_{TY} = \omega_{YT} = 0$), we have
\begin{equation}
(\gamma + \omega_{Y})[(\gamma + \omega_{\nu})(\gamma + \omega_{T}) -
\omega_{g}^{2}] = 0.
\end{equation}
The first root, $\gamma_{1} = - \omega_{Y}$, describes a stable 
diffusive mode. Two other roots correspond to the ordinary buoyancy 
modes one of which can be unstable if $\omega_{g}^{2} - \omega_{\nu}
\omega_{T}> 0$. Then, the necessary condition of instability reduces 
to the Schwarzschild condition $(\Delta\nabla T)_z>0$, or in terms 
of the entropy gradient, $ds/dz < 0$.

If $dY/dz \neq 0$ but dissipative effects are negligible
($a_{2}=0$, $a_{0}=0$), we have
\begin{equation}
\gamma^3 + a_1 \gamma = 0
\end{equation}
with $a_{1}= -(\omega_{g}^{2} + \omega_{L}^{2})$. 
The first root is degenerate in this case and the 
two other roots are given by $\pm \sqrt{-a_{1}}$. 
The condition of instability is $(- a_{1}) = \omega_{g}^{2} + 
\omega_{L}^{2} > 0$, or 
\begin{equation}
\left(\frac{\Delta\nabla T}{T}\right)_z - \frac{\delta}{\beta}\frac{dY}{dz}>0,
\end{equation}
that represents the familiar Ledoux criterion.

\section{Instability criteria in the general case.}

Generally, when $dY/dz \neq 0$ and dissipative effects are important,
the conditions of instability are more complex and depend on a horizontal 
wavevector of perturbations, $k$. The temperature and lepton
gradients required for instability can be quite different for 
perturbations with different $k$ and to obtain the criteria we have to 
find the minimal values of these gradients. From the properties of
kinetic coefficients we have $\kappa_{T} \lambda_{Y} - \kappa_{Y}
\lambda_{T} > 0$, therefore we can deduce 
from conditions (\ref{3cond}) that the gradients are minimal for 
perturbations with the wavevector $k$ minimizing the quantity $Q^{6}/k^{2}$,
which is reached at $k^{2} = (\pi/d)^{2}/2$. Then, we have 
\begin{equation}
\left( \frac{Q^{6}}{k^{2}} \right)_{min} = \frac{27}{4} \left( 
\frac{\pi}{d} \right)^{4}.
\end{equation}
The instability criteria can now be obtained from 
the last two conditions in  (\ref{3cond}), which read, respectively,
\begin{eqnarray}
\frac{dY}{dz} (\delta \kappa_{T} - \beta \kappa_{Y} ) &-& 
\left(\frac{\Delta\nabla T}{T}\right)_z (\beta \lambda_{Y} - \delta 
\lambda_{T} )    
\nonumber \\
&+& \frac{27 \pi^{4} \nu}{4 g d^{4}} (\kappa_{T} 
\lambda_{Y} - \kappa_{Y} \lambda_{T}) < 0 \, ,
\label{c1}
\\
\frac{d Y}{dz} (\delta ( \nu + \lambda_{Y} ) +  \beta \kappa_{Y}) 
&-&\left(\frac{\Delta\nabla T}{T}\right)_z (\beta ( \nu + \kappa_{T} ) + 
\delta \lambda_{T}) +
\nonumber \\
&+& \frac{27 \pi^{4}}{4 g d^{4}} (\kappa_{T} + \lambda_{Y})
\left[ (\nu+\kappa_{T})(\nu+ \lambda_{Y}) - \lambda_{T} \kappa_{Y}\right] < 0 .
\label{c2}
\end{eqnarray}

In the case of a negligible diffusivity, the first criterion yields 
the Rayleigh-Taylor condition of instability, $\delta dY/dz < 0$. This 
instability can be responsible, for instance, for the salt fingers 
phenomena in the terrestrial oceans. Therefore, following Bruenn \& 
Dineva (1996), we can conventionally call the instability associated 
to condition (\ref{c1}) neutron
fingers. We refer to a {\it neutron finger unstable} region as a region where 
condition (\ref{c1}) is fulfilled but not condition (\ref{c2}). In the case 
of a ``standard'' transport with small viscosity and 
diffusivity, the second criterion yields the Schwarzschild criterion of
convection, although convection in PNS is, in general, quite different from 
the Schwarzschild convection and may arise in oscillatory or nonoscillatory 
regimes. A region is said to be {\it convectively unstable} when both 
conditions (\ref{c1}) and (\ref{c2}) are satisfied. If condition (\ref{c2}) is
satisfied but condition (\ref{c1}) is not, the system is said to be 
{\it semiconvectively unstable}.

Despite the fact that convective instabilities in PNS
were originally
argued by applying the Ledoux criterion (see, e.g., Epstein 1979,
Livio, Buchler \& Colgate 1980), it appears that the Ledoux criterion 
is not the valid criterion for instability if we allow for conduction of heat,
diffusion of particles and/or viscosity. In this case, the true criteria of 
instability are (\ref{c1}) and (\ref{c2}), and
only in two limiting cases
our criteria reduces to the Ledoux form.

Generally, the region where any of the conditions (\ref{c1},\ref{c2}) 
are fulfilled can be very different to that given by the Ledoux criterion.

\section{Results and discussion.}

To obtain the different thermodynamical derivatives, diffusion coefficients
and conductivities appearing in the coefficients of the dispersion equation,
we used the results from numerical simulations of PNS evolution performed
by Pons {\it et al.} (1999). The results discussed in this paper correspond
to the model labelled GM3np, with a baryonic mass of 1.6 $M_{\odot}$.
The simulation was carried out using a Henyey--like spherically symmetric
evolution code, coupled to a 1D neutrino transport scheme in the 
flux--limited diffusion approximation. Details about the code and
the calculation of the diffusion coefficients using opacities
consistent with the underlying EOS (Reddy, Prakash \& Lattimer, 1998)
can be found in Pons {\it et al.} (1999). 

Since calculations of viscosity at
high densities are unreliable (compare, for instance, van den Horn 
\& van Weert 1981, Goodwin \& Pethick 1982, Thompson \& Duncan 1993), we 
calculate the 
unstable region for different values of $\nu$. In Figure 1 (upper panel) 
we plot the stability regions of a 1.6 $M_{\odot}$ PNS corresponding to
case $\nu=0$. In all the simulations we use a value for $d$ given by 
the pressure scale, $d=\mid d\ln p/ dr \mid^{-1}$.  
Three different situations can clearly be distinguished. Initially, there 
is a small convectively unstable zone (darkest) near the surface that 
lasts for only a few seconds, surrounded by a neutron finger unstable region
(intermediate gray tone). The inner part of the star is stable (lighter).
After a few seconds, the neutron finger instability moves inward, 
occupying a large portion of the PNS, whereas the stable region shrinks.
Later on, at $t=12$ s, the innermost core becomes convectively unstable
for a few seconds, while the neutron finger unstable region begins
to shrink. By about $t=30$ s, the PNS is mostly stable, except a small
region near the center that is still subject to the neutron fingers 
instability.
Instability completely disappears after $\approx 40$ s.

To study the effect of viscosity, Figure 1 (lower panel)  shows the time
dependence of the boundaries of unstable regions for $\nu = \eta^{2} 
\kappa_{T}$ (van den Horn \& van Weert 1981). Qualitatively, the
unstable regions evolve in a similar fashion to the inviscid case.
The reason why viscosity hardly influences 
the neutron finger instability region is that the third term on the 
l.h.s. of (\ref{c1}), which 
describes dissipation of perturbation due to neutrino transport
and depends on $\nu$, is relatively small compared to the first
two terms which are independent of $\nu$. Therefore, the neutron finger 
unstable region is approximately the same as in Fig.1 at any 
evolutionary stage. On the contrary, convective instability is
sensitive to the value of viscosity through condition (\ref{c2}). The third 
term on the l.h.s.
of (\ref{c2}) again does not yield too much contribution, however, two other
terms depend on viscosity. Due to this, the region unstable to 
convection turns out to be more extended at high viscosity.
Notice that in the case $\eta \gg 1$ the 
viscosity is much greater than the diffusivity and thermal diffusivity but, 
nevertheless, is not sufficiently large to make the term in the r.h.s. 
important, thus recovering Ledoux criterion from (\ref{c2}).
Note, however, that for both models the
convectively unstable region always lies inside the region unstable
to neutron fingers. In our simulations we have not found any region unstable
to semiconvection. Although this result might seem different from Bruenn 
\& Dineva (1996) 
we should keep in mind that the thermodynamic conditions used in this work to 
describe the interior of the PNS are 
quite different from the conditions used by Bruenn \& Dineva (1996)
to describe the matter just below the neutrinosphere of the collapsed core.
Bruenn \& Dineva (1996) studied the instability in a region, typically  with a
density of 3$\times 10^{13}$ g/cm$^3$and with a high entropy per baryon 
$s=4$, since at
such early times ($\approx 50-230$ ms after bounce) the excess of entropy of 
the shocked mantle has not been yet radiated away. In this work, however, we 
focus on the long term evolution of a PNS, and most of the 
region of our interest is at supranuclear densities ($\rho > 3\times 10^{14}$
g/cm$^3$) and moderate entropies.

Since the superadiabatic temperature gradient and the gradient of lepton 
number play a fundamental role in the stability criterion we have shown in
Figure 2 graphics of these quantities for different times in the evolution. 
We plot thicker lines for the unstable regions according to the model shown 
in the lower panel of Figure 1.
As we can see from the figures, at early and middle times in the
evolution, instability region corresponds approximately to that with a
positive value of the superadiabatic temperature gradient.
The effect of the negative lepton fraction gradient and/or dissipative
processes increase only slightly the unstable region.
At late time however, the unstable region almost disappears while the
superadiabatic temperature gradient remains positive in a much wider region.
This is chiefly caused by the change in the sign of the
thermodynamic derivative $\delta$, when the lepton fraction drops some critical
value.

In the upper panel of Figure 3 we show the maximum of the real parts of the 
roots
of equation (\ref {cubic}) at two different stages in the evolution. 
A positive value of this quantity
implies instability and, in that case, the instability growth time is 
calculated  by taking its inverse (lower panel).
As we can contrast in the lower panel of Figure 3 the characteristic
growth time of instabilities at the beginning of the evolution (t=1 s) is of 
the order of 
tenths of millisecond. Since this time is much shorter 
than the
evolution time scale, convective heat transport might be important in this 
region.
At the middle stages of the evolution (t=20 s), two different unstable
regions can be distinguished, a first one with a growth time of the
order of 1 ms (convection) and a second region with associated 
growth times of the order of few tens of millisecond (corresponding to the 
region unstable to neutron fingers).
The growth time in both cases is, again, 
much shorter than the time evolution and, in principle,
instabilities can grow fast enough to influence the heat transport. 

In summary, we have to emphasize that neutrino transport plays
a significant role in the stability properties of PNSs. 
Criteria (25) and (26) obtained in the present paper correspond
properly to the conditions of PNS where heat transport and
diffusion can be sufficiently fast. These criteria differ essentially 
from those used by other authors in this context.
Our analysis shows that the unstable
region inside the PNS can be more extended than predicted, for
instance, by the Ledoux criterion alone. Generally, neutron fingers 
instability arises in a more extended region of a stellar volume and 
lasts a longer time than convection. Therefore, this instability
seems to be as important as convection for the early evolution of a newly 
born neutron star. 
Our main conclusion is that, surprisingly, diffusive 
effects allow for the existence of new unstable modes and the role
of convection in PNS should be carefully reexamined.

\section*{Acknowledgement}
This research was supported in part by the Spanish Ministerio of Educaci\'on
y Cultura (grant PB97-1432) and the Russian Foundation of Basic Research
(grant 00-02-04011). V.U. thanks the University of Valencia for financial
support and hospitality.

\section*{References}

\noindent
Aleksandrov A.D., Kolmogorov A.N., Laurentiev M.A. 1963. Mathematics: Its
Content, Methods, and Meaning. MIT press.\\
Arnett W. D. 1987, ApJ, 319, 136 \\
Bruenn S., Dineva, T. 1996, ApJ, 458, L71 \\
Bruenn S., Mezzacappa A. 1994, ApJ, 433, L45 \\
Bruenn S., Mezzacappa A., Dineva T. 1995, Phys. Rep., 256, 69  \\
Burrows A., Fryxell B. 1993, ApJ, 418, L33 \\
Burrows A., Lattimer J. 1986, ApJ,  307, 178 \\
Burrows A., Lattimer J. 1988, Phys. Rep., 163, 151 \\
Colgate S., Petschek A.1980, ApJ, 238, L115 \\
Thompson C., Duncan R. 1993, ApJ, 408, 194 \\
Epstein R. 1979, MNRAS, 188, 305 \\
Goodwin B., Pethick C. 1982, ApJ, 253, 812 \\
Grossman, S.A., Narayan, R., Arnett, D., 1993, ApJ, 407, 284 \\
Hillebrandt W. 1987, in High Energy Phenomena around Collapsed Stars
(ed. F.Pacini), Dordrecht, Reidel, 73 \\
Imshennik V.S., Nadezhin D.K. 1972, Sov. Phys. JETP, 36, 821 \\
Janka H.-T., Muller E. 1994, A\&A, 290, 460 \\
Keil W., Janka H.-T. 1995, A\&A, 296, 145 \\
Keil W., Janka H.-T., Muller E. 1996, ApJ, 473, L111 \\
Landau L., Lifshitz E. 1959, Fluid Mechanics, London, Pergamon \\
Lattimer, J.M., \& Mazurek, T.J., 1981, ApJ, 246, 955 \\
Livio M., Buchler J., Colgate S. 1980, ApJ, 238, L139 \\
Mezzacappa, A., Calder, A.C., Bruenn, S.W., Blondin, J.M., Guidry, M.W. 
Strayer, M.R., and Umar, A.S., 1998, ApJ, 493, 848 \\
Pons J.A., Reddy S., Prakash M., Lattimer J., Miralles J.A. 1999, ApJ, 
513, 780 \\
Reddy S., Prakash M., Lattimer J., 1998, Phys. Rev., D58, 013009. \\
Smarr L., Wilson J., Barton R., Bowers R. 1980, ApJ, 246, 515 \\
Sumiyoshi K., Suzuki H., Toki H. 1995, A\&A, 303, 475 \\
van den Horn L., van Weert C. 1981, ApJ, 251, L97 \\ 

\section*{Figure captions}

\vspace*{1cm}
\noindent Fig. 1.-- Evolution of the different unstable zones in a $M=1.6 M_{\odot}$
PNS. The darker zone corresponds to the convectively unstable regions,
the lighter zone to convectively stable regions and the intermediate
zone presents the secular instability usually denoted by neutron fingers.
The upper and lower panels displays the results of the inviscid and
viscous ($\nu =\eta^2 \kappa_T$) models, respectively.

\vspace*{1cm}
\noindent Fig. 2.-- Superadiabatic temperature gradient (left panels) and
lepton fraction gradient (right panels) for three different times in the
evolution. Thicker lines correspond to the unstable region according to
Figure 1.

\vspace*{1cm}
\noindent Fig. 3.-- Maximum value of the real parts of the roots of equation 
(17) at two different stages in the evolution (upper panel) and  growth time 
for the unstable regions (lower panel).

\begin{figure}
\begin{center}
\epsfxsize=6.in
\epsfysize=7.in
\epsffile{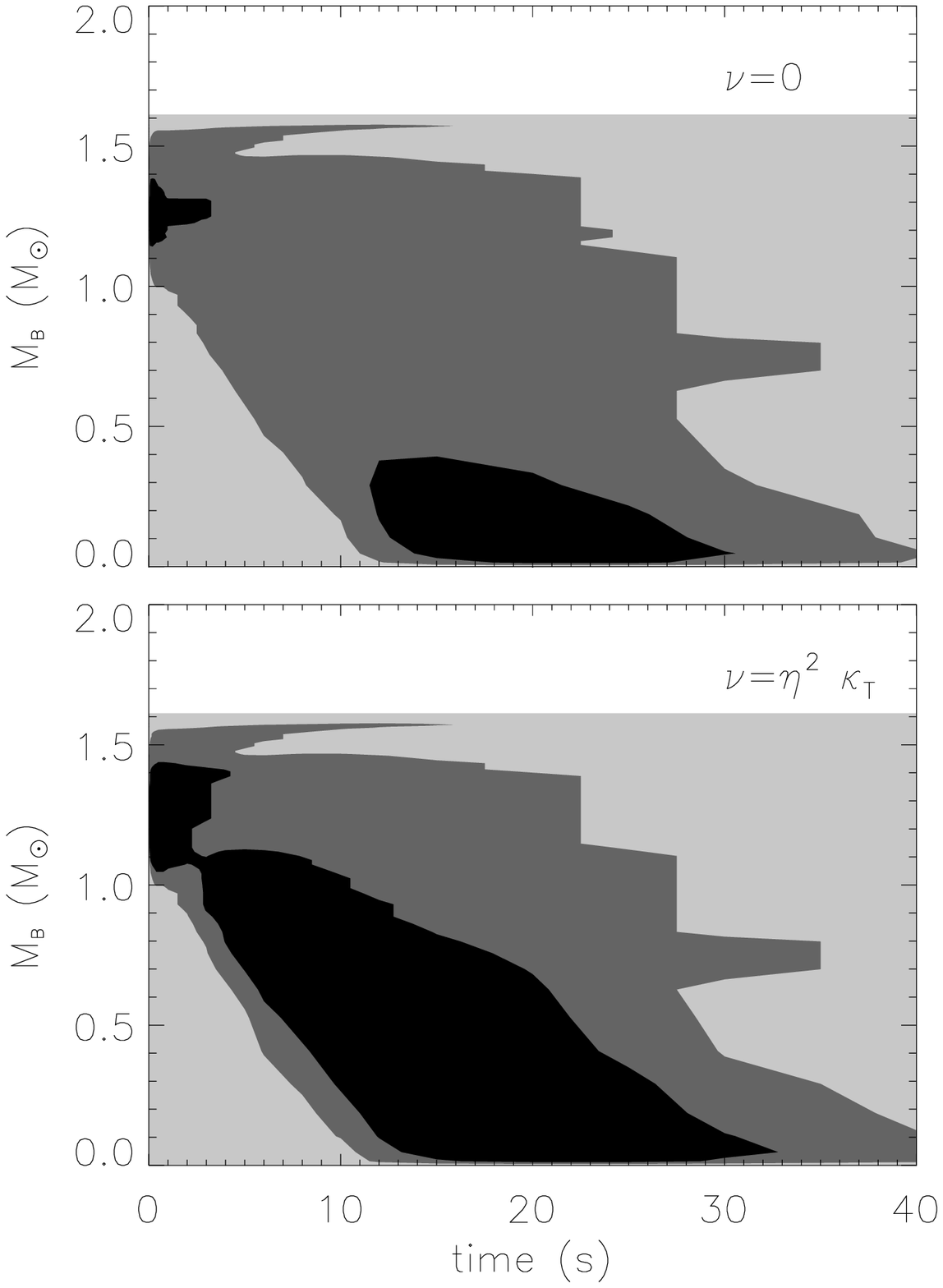}
\end{center}
\caption{}
\end{figure}

\begin{figure}
\begin{center}
\epsfxsize=6.in
\epsfysize=7.in
\epsffile{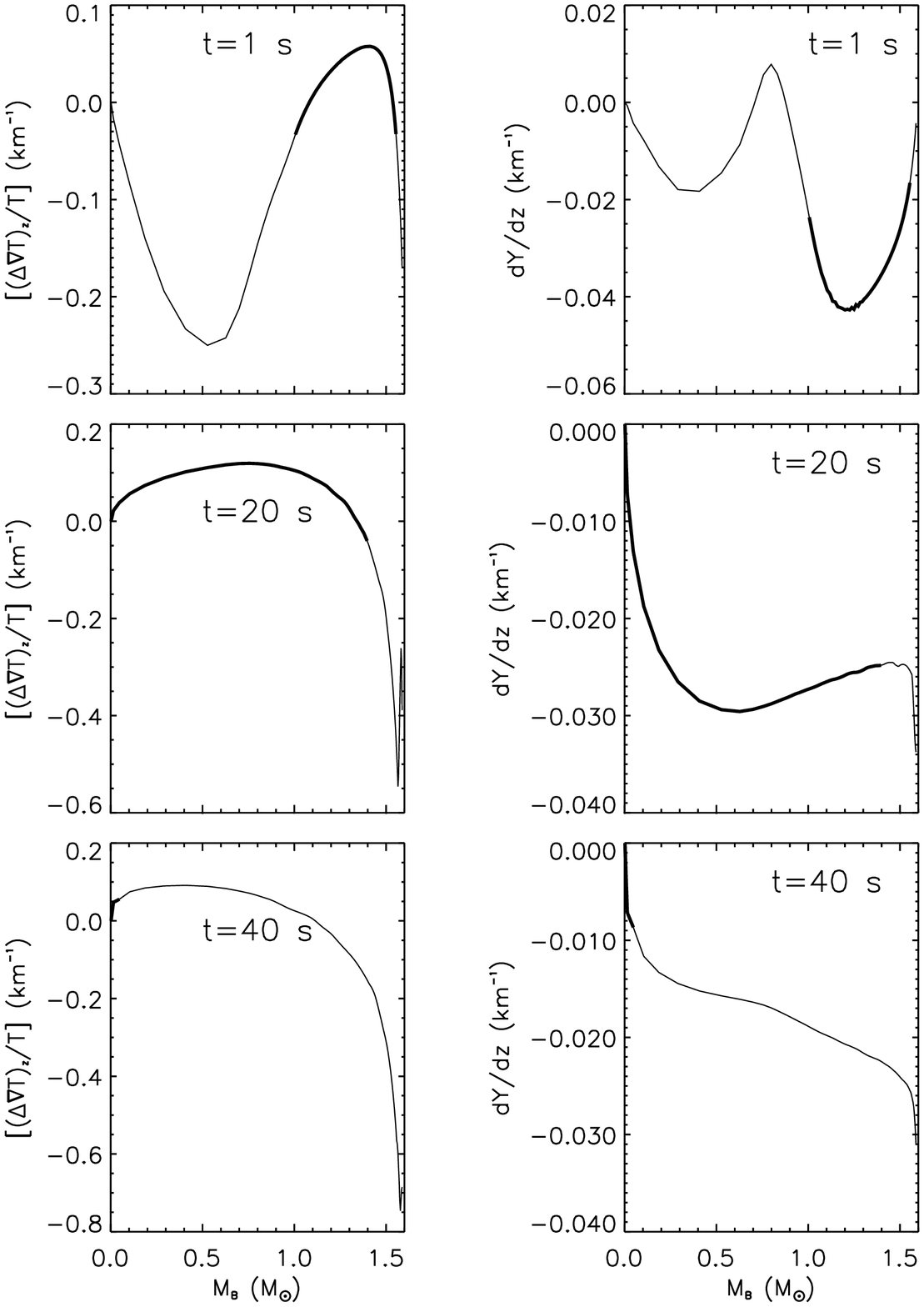}
\end{center}
\caption{}
\end{figure}

\begin{figure}
\begin{center}
\epsfxsize=6.in
\epsfysize=7.in
\epsffile{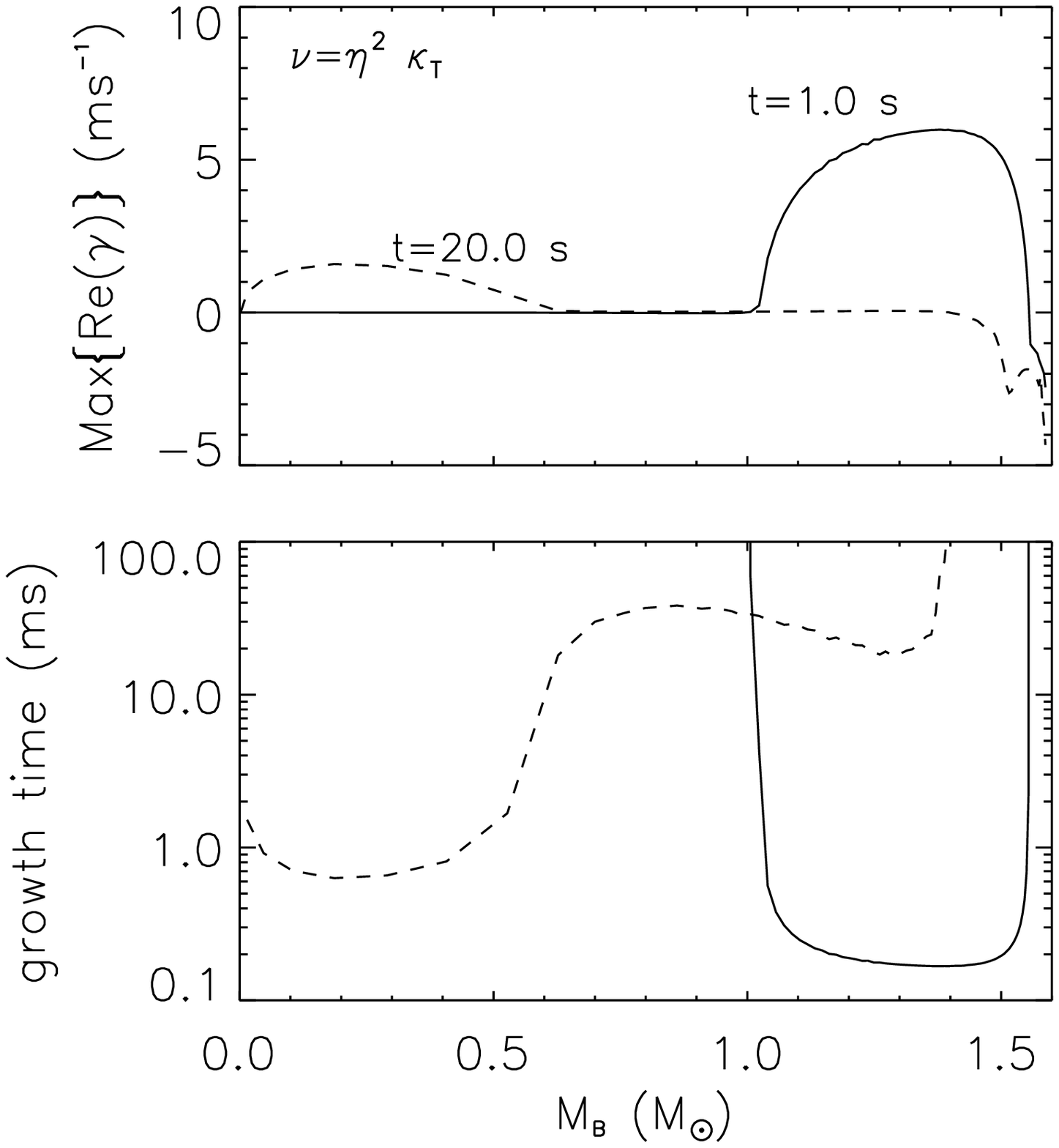}
\end{center}
\caption{}
\end{figure}

\end{document}